\documentclass[preprint,12pt]{elsarticle}

\usepackage{epstopdf}

\usepackage{mathtools}
\usepackage{amsmath}

\begin{document}

\begin{frontmatter}



\title{Gouy phase of type-I SPDC-generated biphotons}

\author[label1,label2]{F. C. V. de Brito}
\address[label1]{Departamento de F\'{\i}sica, Universidade Federal
do ABC, S\~ao Paulo, SP, Brazil}
\address[label2]{Departamento de F\'{\i}sica, Universidade Federal do
Piau\'{\i}, Campus Ministro Petr\^{o}nio Portela, CEP 64049-550,
Teresina, PI, Brazil}

\author[label2,label6]{I. G. da Paz}

\author[label3]{Brigitte Hiller}
\address[label3]{ CFisUC - Department of Physics, University of
Coimbra, 3004-516 Coimbra, Portugal}

\author[label4]{Jonas B. Araujo}
\address[label4]{Departamento de F\'{\i}sica Matem\'atica,
Instituto de F\'{\i}sica, Universidade de S\~ao Paulo, C.P. 66.318,
S\~ao Paulo - SP, 05315-970, Brazil}

\author[label1]{Marcos Sampaio}

\address[label6]{Corresponding author\\
e-mail: irismarpaz@ufpi.edu.br}

\begin{abstract}
We consider a double Gaussian approximation to describe the
wavefunction of twin photons (also called a biphoton) created in a
nonlinear crystal via a type-I spontaneous parametric downconversion
(SPDC) process. We find that the wavefunction develops a Gouy phase
while it propagates, being dependent of the two-photon correlation
through the Rayleigh length. We evaluate the covariance matrix and
show that the logarithmic negativity, useful in quantifying
entanglement in Gaussian states, although Rayleigh-dependent, does
not depend on the propagation distance. In addition, we show that
the two-photon entanglement can be connected to the biphoton Gouy
phase as these quantities are Rayleigh-length-related. Then, we
focus the double Gaussian biphoton wavefunction using a thin lens
and calculate a Gouy phase that is in reasonable agreement with the
experimental data of D. Kawase \emph{et al.} published in Ref. [1].
\end{abstract}



\end{frontmatter}

\section{Introduction}

Since its first detection in 1890 by L. G. Gouy \cite{gouy1,gouy2},
the Gouy phase and its properties have been extensively studied
[4-11]. This phase appears whenever a wave is constrained
transversally to its propagation, which includes diffraction through
slits and focus by lenses. The acquired phase depends on the type of
transversal confinement and on the geometry of the waves. For
example: line-focusing a cylindrical wave propagating from $-\infty$
to $+\infty$ yields a Gouy phase of $\pi/2$, while point-focusing a
spherical wave in the same interval yields a Gouy phase of $\pi$
\cite{feng2001}; Gaussian matter wave packets diffracting through
small apertures pick up a Gouy phase of $\pi/4$ \cite{Paz4}.

The Gouy phase has been detected in various scenarios, including
acoustic and water waves \cite{cond,elec2,elec1}, surface
plasmon-polaritons with non Gaussian spatial properties \cite{zhu},
focused cylindrical phonon-polariton wave packets in LiTaO$_3$
crystals, and more recently for electron waves
\cite{cond,elec2,elec1}. Its presence in many systems justifies
potential applications. To name a few, the Gouy phase is fundamental
in evaluating the resonant frequencies in laser cavities
\cite{siegman}, in phase-matching in strong-field and high-order
harmonic generation \cite{Balcou}, and in describing the spatial
profile of laser pulses with high repetition rate \cite{Lindner}. In
addition, an extra Gouy phase appears in optical and matter waves
depending on the orbital angular momentum's magnitude
\cite{Allen,elec2}. In a recent work, it was found that the Gouy
phase may cause nonlocal effects that modify the symmetries of
self-organization in atomic systems \cite{Guo}. This phase may also
be useful in communication and optical tweezers using structured
light \cite{Khoury}.

The Gouy phase is also relevant in coherent matter waves, as shown
for the first time in \cite{Paz4,Paz1,Paz2,Ducharme}. Following
that, experiments were performed in a number of systems, including
Bose-Einstein condensates \cite{cond}, electron vortex beams
\cite{elec2} and astigmatic electron waves \cite{elec1}. Gouy phases
in matter waves also display potential applications, namely: they
can be used in mode converters in quantum information systems
\cite{Paz1}, in the generation of singular electron optics
\cite{elec1} and in the study of non-classical (exotic or looped)
paths in interference experiments \cite{Paz3}. In this work, we are
interested in the Gouy phase of entangled photon pairs generated in
a type-I SPDC process.

An SPDC process generates a pair of entangled photons respecting
energy-momentum conservation. These processes happen with extremely
low probability -- around $10^{-7}$ \cite{BookSPDC}. Because the
first experiments involved non degenerate emerging photon beams, one
with frequency in the IR and the other in the visible range, they
were named idler and signal, respectively \cite{Yanhua}. The
emerging photons in these processes are highly correlated in energy,
momentum, polarization and angular momentum \cite{ENTANGBIPH}. They
emerge after a pump beam, with frequency $\omega_p$, goes through a
nonlinear crystal, generating (in those very rare cases) two lower
energy photons, the idler and signal, with frequencies $\omega_i$
and $\omega_s$. The type of SPDC depends on the polarization of the
emerging photons with respect to the incoming pump beam. For
example, in a type-I SPDC, the signal and idler photons display
parallel polarizations, both orthogonal to the pump beam's, and form
a cone aligned to the pump beam's direction. In a type-II SPDC the
signal and idler photons have orthogonal polarizations and emerge in
2 different cones. The spatial distribution of the emerging beams is
a consequence of energy-momentum conservation: $\omega_p = \omega_i
+ \omega_s $ and $\vec{k}_p = \vec{k}_i  + \vec{k}_s $. This also
causes the high degree of energy-momentum correlations between the
emerging beams. For a more details, please consult Ref.
\cite{OURPRA} and the references therein. In fact, it is possible to
control the correlations between different degrees of freedom in the
generated pairs \cite{Menzel2007}. In this work, we will consider
twin photons with wavelength $702.2 \,$nm, typically used in
interferometry experiments, such as in Refs.
\cite{Monken1999,Nature2017}.

Regarding the entanglement, the Schmidt number plays an important
role. The propagation dynamics of spatially entangled biphotons was
explored via the Schmidt number in Ref. \cite{Matthew}. Like the
logarithmic negativity, the Schmidt number is
propagation-distance-independent and the entanglement migrates
between amplitude and transverse phase. In this work we will explore
the two-photon entanglement by means of the longitudinal Gouy phase
of the double Gaussian approximation for the biphoton wavefunction.
We calculate the Gouy phase for this approximated biphoton
wavefunction and show that it is related with the photon correlation
generated in the nonlinear crystal in a type-I SPDC process. Even
though the photon entanglement is time-independent, whereas the Gouy
phase is time dependent, these quantities become related by the
Rayleigh length. More interestingly we show that the approximated
biphoton Gouy phase fits well the experimental data published in
Ref. \cite{Kawase}.

The article is organized as follows: in section II we propagate the
double Gaussian biphoton wavefunction and obtain the corresponding
Gouy phase. In section III we evaluate the covariance matrix and the
logarithmic negativity and show that the two-photon entanglement is
longitudinal-distance-independent. We observe that the entanglement
measured by the logarithmic negativity and the Gouy phase are
related by the Rayleigh length. In section IV we focus the double
Gaussian biphoton wavefunction and use the corresponding Gouy phase
to analyze the existing experimental data. In section V we draw our
concluding remarks.

\section{Propagation of biphoton wavefunction and Gouy phase}
In this section we propagate a double Gaussian biphoton wavefunction
using free-particle propagators. Then, we obtain a
Gaussian solution expressed in terms of real terms and phases. One
of the phases is the Gouy phase, which is transverse-position-independent and is a function of the longitudinal distance of
propagation, the beam pump parameters, and the twin photon
correlation. We consider as the initial biphoton wavefunction the
following entangled state \cite{ford,dorlas,bp}
\begin{equation}
\Psi(x_1, x_2)=\frac{1}{\sqrt{\pi \sigma \Omega}}
e^{\frac{-(x_1-x_2)^2}{4\sigma^2}}
e^{\frac{-(x_1+x_2)^2}{4\Omega^2}}, \label{estadoinicial}
\end{equation}
which is the generalized EPR state for the momentum-entangled
particles. Here, $\Omega$ and $\sigma$ quantify the position and
momentum uncertainties of the packet, i.e., $\Delta x_1=\Delta
x_2=\sqrt{\Omega^2+\sigma^2}$ and $\Delta p_{x1}=\Delta
p_{x2}=(\hbar/4)\sqrt{(1/\Omega^2)+(1/\sigma^2)}$. This approximated
biphoton state is correlated only if $\Omega\neq\sigma$ and
$\Omega=\sigma$ corresponds to a non entangled state which factors
as a product of two Gaussians \cite{bp}.

We will work with relative coordinates $r=(x_1+x_2)/2$ and
$q=(x_1-x_2)/2$ since these are convenient for
calculations. Thus, the initial wavefunction that represents the
entangled state can be rewritten as
\begin{equation}
 \Psi (r,q)=\frac{1}{\sqrt{\pi \sigma \Omega}} e^{-\frac{q^2}{\sigma^2}} e^{-\frac{r^2}{\Omega^2}}.
\label{estadoinicial}
\end{equation}
The state describing the biphoton free propagation can be written as
\begin{equation}
 \Psi (r,q,t) =\int_{r', q'} K_r(r,t;r',0) K_q(q,t;q',0) \psi(r',q'),
 \end{equation}
where the propagation kernels of a longitudinal distance $z=ct$ for
the two photons are given by

\footnotesize
\begin{equation}\label{propagador}
\begin{split}
K_r(r,r',z)=\sqrt{\frac{1}{i\lambda
z}}\exp{\bigg[-\frac{2\pi(r-r')^2}{i\lambda z}\bigg]}, \\
K_q(q,q',z)=\sqrt{\frac{1}{i\lambda
z}}\exp{\bigg[-\frac{2\pi(q-q')^2}{i\lambda z}\bigg]}.
\end{split}
\end{equation}

\normalsize The state after a general distance $z$ can be evaluated
as
\begin{equation}\label{psifp}
\Psi(r,q,z)= \frac{1}{\sqrt{4\pi w_{+}(z)w_{-}(z)}}\exp \bigg
\lbrace -\bigg[\frac{ r^2}{w_{+}^{2}(z)} +\frac{q^2}{w_{-}^2(z)}
\bigg] \bigg \rbrace \\ \times \exp \bigg \lbrace -i
\bigg[-\frac{k_0}{r_{+}}r^2-\frac{k_0}{r_{-}}q^2+\zeta(z)\bigg]
\bigg \rbrace,
\end{equation}
where,
\begin{equation}\label{b}
w_{\pm}^2(z)= \Omega^2
\bigg[1+\bigg(\frac{z}{z_{0\pm}}\bigg)^2\bigg],\hspace{0.5cm}
r_{\pm}(z)=z \bigg[1+\bigg(\frac{z_{0\pm}}{z}\bigg)^2\bigg],
\end{equation}

\begin{equation}\label{r}
z_{0+}=k_0\Omega^2, \hspace{0.2cm} z_{0-}=k_0\sigma^2 \hspace{0.2cm}
\text{and} \hspace{0.2cm} k_0=2\pi/\lambda.
\end{equation}

Now, considering the analogy with the classical Gaussian laser beam
we can identify the biphoton wavefunction terms as: $w_{\pm}(z)$ is
the beam width, $r_{\pm}(z)$ the radius of curvature of the wave
fronts and $z_{0\pm}$ the corresponding Rayleigh lengths. The
function $\zeta(z)$ is the biphoton Gouy phase that, after some
algebraic manipulations, is written as
\begin{eqnarray}\label{mu1mu2}
\zeta(z)&=&\frac{\zeta_{+}(z)}{2}+\frac{\zeta_{-}(z)}{2}\nonumber\\
&=& \frac{1}{2} \arctan \bigg[z \bigg(
\frac{z_{0+}+z_{0-}}{z_{0+}z_{0-}-z^2} \bigg)\bigg],
\end{eqnarray}
where $\zeta_{+}(z)=\arctan(z/z_{0+})$ and
$\zeta_{-}(z)=\arctan(z/z_{0-})$. We can see that this phase is
propagation-distance-dependent. It carries the properties of the
laser pump beam and the nonlinear crystal through the parameter
$\sigma=\sqrt{\frac{L_p \lambda_p}{6 \pi}}$, where $\lambda_p$ is
the laser pump wavelength  and $L_p$ the crystal length. The
two-photon correlation dependence can be measured through the
parameter $\Omega$.

\begin{figure}[htp]
\centering
\includegraphics[height=4cm,width=5 cm]{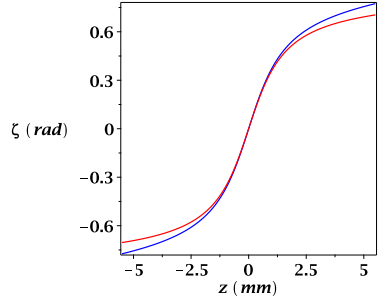}
\caption{Biphoton Gouy phase as a function of $z$. The curve in blue
corresponds to $\Omega=5\sigma$, and the curve in red corresponds to
$\Omega=10\sigma$. \label{gouy_z}}
\end{figure}

In Figure \ref{gouy_z} we show the plot of the biphoton Gouy phase
as a function of $z$. As in Ref. \cite{brida} we consider the
following set of parameters: biphoton wavelength
$\lambda=702\;\mathrm{nm}$, laser pump wavelength
$\lambda_p=351.1\;\mathrm{nm}$ and the crystal length
$L_p=7.0\;\mathrm{mm}$. This enables us to obtain
$\sigma=\sqrt{\frac{L_p \lambda_p}{6 \pi}}=11.4\;\mathrm{\mu m}$ and
$z_{0-}=k_0\sigma^{2}=1.2\;\mathrm{mm}$. For the curve in blue we
consider $\Omega=5\sigma$ and for the red curve we consider
$\Omega=10\sigma$.
As we can observe, the maximum variation of the Gouy phase is $\pi/2$,
characterizing one-dimensional free propagation from
$z=-\infty$ to $z=+\infty$ with the beam waist located at the origin
$z=0$ at the position of the crystal. Also, the smaller correlation
produces a larger Gouy phase variation as we can see by comparing
the curves in blue and red.

\section{Entanglement and Gouy phase}
Here we show how the two-photon entanglement is related with the
parameters $\sigma$ and $\Omega$ and therefore with the Rayleigh
length $z_{0\pm}$. In fact, the Rayleigh-length-dependence
establishes a connection between the Gouy phase and two-photon
entanglement. A good measure of entanglement for Gaussian states is
the logarithmic negativity which is calculated through the
covariance matrix. In the symplectic form the covariance matrix can
be written as \cite{ford,vidal}
\begin{eqnarray}\label{covariance}
M = \left[
\begin{array}{cccc}
g & 0 & c & 0 \\
0 & g & 0 & c' \\
c & 0 & h & 0 \\
0 & c' & 0 & h
\end{array}
\right]
\end{eqnarray}
which is related to \\ 

\hspace{-1cm}
\(G=
\begin{bmatrix}
 \frac{\langle x_1^2\rangle}{L^2}  & \frac{\langle x_1 p_1 +p_1x_1 \rangle}{2 \hbar}  \\
\frac{\langle x_1p_1 +p_1x_1 \rangle}{2 \hbar} & \frac{L^2\langle
p_1^2 \rangle}{\hbar^2} \nonumber
\end{bmatrix},
 H=
\begin{bmatrix}
 \frac{\langle x_2^2\rangle}{L^2}  & \frac{\langle x_2p_2 +p_2x_2 \rangle}{2 \hbar}  \\
\frac{\langle x_2p_2 +p_2x_2 \rangle}{2 \hbar} & \frac{L^2\langle
p_2^2 \rangle}{\hbar^2} \nonumber
\end{bmatrix},
\; C=
\begin{bmatrix}
 \frac{\langle x_1 x_2 \rangle}{L^2}  & \frac{\langle x_1p_2  \rangle}{ \hbar}  \\
\frac{\langle x_2p_1 \rangle}{\hbar} & \frac{L^2\langle p_1 p_2
\rangle}{\hbar^2} \nonumber
\end{bmatrix}
\),

\vspace{0.5cm} \normalsize \noindent through the simple relations
$\det G= g^2$, $\det H= h^2$ and $\det C= cc'$. The constants
$\hbar$ and $L$, which appear in the above matrices, are inserted to
make the matrix $M$ dimensionless. For the next calculations $L$ can
be disregarded (see \cite{ford} for further discussion on these
constants). We obtain the quantities of $M$ in Eq.
(\ref{covariance}) as follows
\begin{footnotesize}
\begin{equation}
\langle x_1^2\rangle = \langle x_2^2 \rangle=
(\sigma^2+\Omega^2)\bigg[ \bigg(\frac{z}{ z_{0
+}}\bigg)\bigg(\frac{z}{z _{0 -}}\bigg)+1\bigg],
\end{equation}

\begin{equation}
\langle x_1x_2\rangle =\langle x_2^2 \rangle=
(\sigma^2-\Omega^2)\bigg[ \bigg(\frac{z}{ z_{0
+}}\bigg)\bigg(\frac{z}{z _{0 -}}\bigg)-1\bigg],
\end{equation}

\begin{equation}
\langle p_1^2 \rangle = \langle p_2^2 \rangle= \frac{1}{4} \bigg[
\frac{1}{\Omega^2}+\frac{1}{\sigma^2}\bigg], \hspace{0.5cm} \langle
p_1p_2\rangle = \frac{1}{4} \bigg[
\frac{1}{\Omega^2}-\frac{1}{\sigma^2} \bigg],
\end{equation}

\begin{equation}
\langle x_1 p_2 \rangle = \langle x_2 p_1 \rangle=\frac{\pi
\hbar}{2}\bigg(\frac{z}{ z_{0 +}}-\frac{z}{z _{0 -}}\bigg),
\end{equation}
and
\begin{eqnarray}
\sigma_{xp}&=&\frac{\langle x_1p_1 +p_1x_1 \rangle}{2}
=\frac{\langle x_2p_2 +p_2x_2 \rangle}{2}\nonumber\\
&=&\frac{\pi \hbar}{2}\bigg(\frac{z}{ z_{0 +}}+\frac{z}{z _{0
-}}\bigg)=\frac{\pi \hbar}{2}[\tan(\zeta_{+})+\tan(\zeta_{-})],
\end{eqnarray}
\end{footnotesize}
where $\zeta_{+}$ and $\zeta_{-}$ are parts of the biphoton Gouy
phase from Eq. (\ref{mu1mu2}). A relation between these two
quantities was obtained previously in the context of a single
particle \cite{Paz1}. Here, we are showing that the biphoton Gouy
phase is part of the logarithmic negativity (entanglement) through
the position momentum covariance.

\begin{figure}[htp]
\centering
\includegraphics[height=4cm,width=6.0 cm]{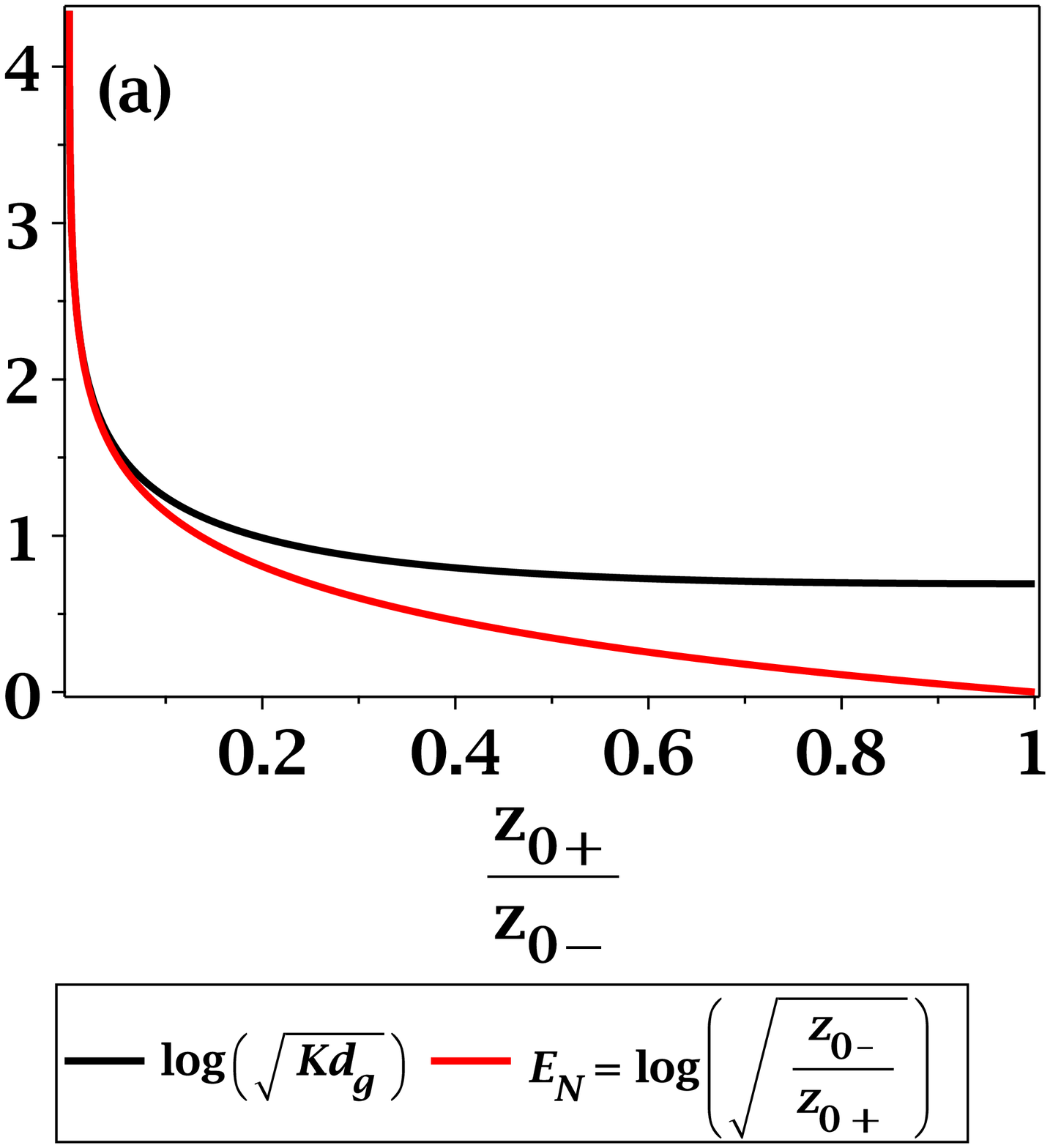}
\includegraphics[height=4cm,width=6.0 cm]{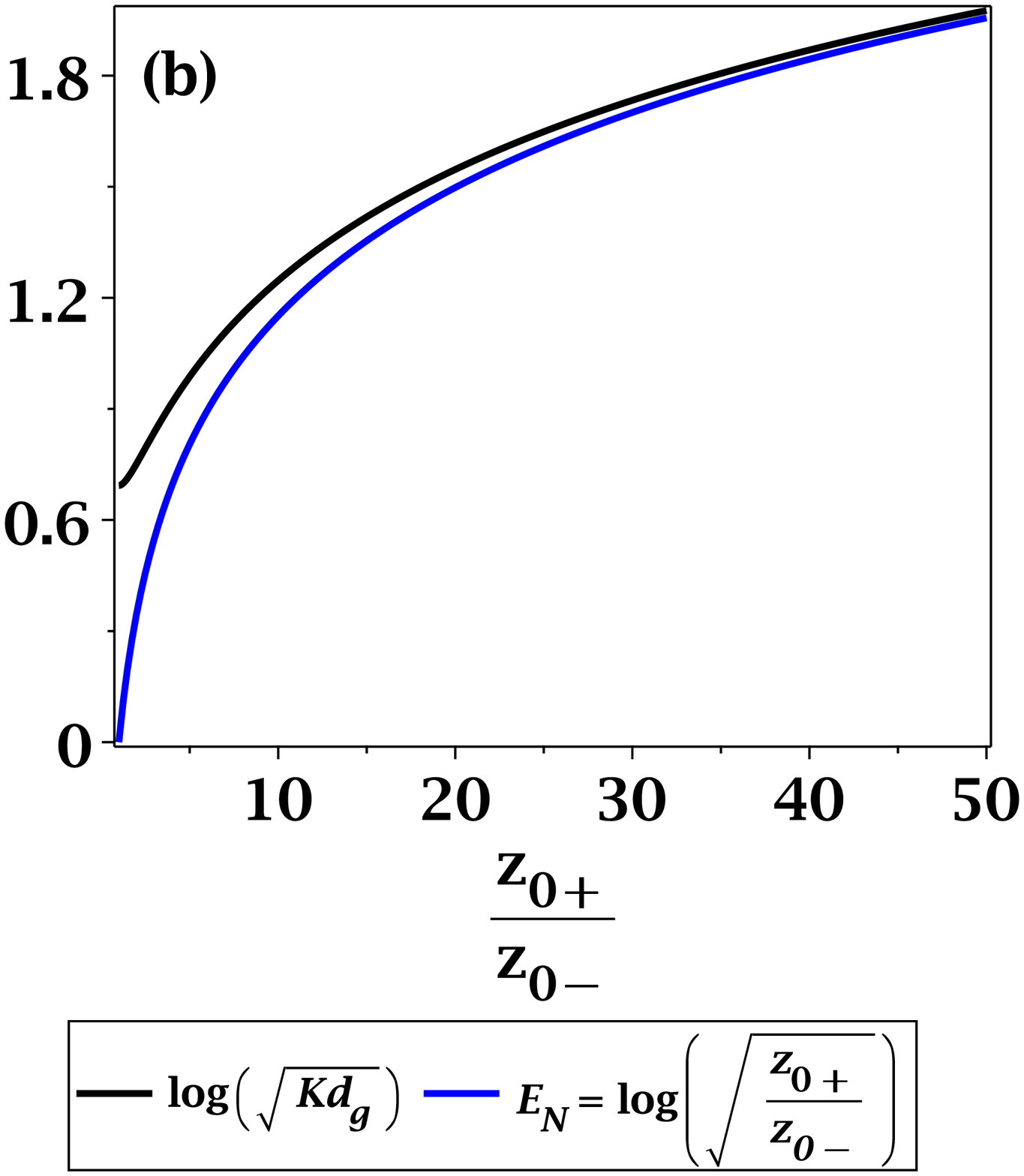}
\caption{(a) Logarithmic negativity and the logarithm of the root
square of the Schmidt number as a function of $z_{0+}/z_{0-}\leq1$
and (b) the same quantities for $z_{0+}/z_{0-}\geq1$. These
quantities agree in the limits $z_{0+}/z_{0-}\ll1$ and
$z_{0+}/z_{0-}\gg1$.\label{schmidtfig}}
\end{figure}

A strong necessary condition for an entanglement quantifier is that
it has to be zero if the state is separable. The Peres-Horodecki
criterion says that if a state is separable, the transpose partial
matrix of the state has a non-negative spectrum. In that context,
the Gaussian state is separable if and only if the minimum value of
the symplectic spectrum of $M^{T_2}$ is greater than $1/2$ (the
lowest value allowed by the uncertainty principle). Thus, a good
measure of entanglement for all Gaussian states is the logarithmic
negativity \cite{ford,vidal}
\begin{equation}\label{negatividade}
E_N=max \lbrace0,-\log(2 \nu_{min})\rbrace,
\end{equation}
where,  $\nu_{min}$ is the lowest symplectic eigenvalue of
$M^{T_2}$. The equation determining the symplectic eigenvalues is
$\nu^4+(g^2+c^2-2 cc')\nu^2+det (M)=0$, with solutions $\pm
i\nu_{\alpha}$, $\alpha=1,2$ where $\nu_{\alpha}$ is the symplectic
spectrum. Therefore, $\nu_1= (\Omega/2\sigma)$ and
$\nu_2=(\sigma/2\Omega)$. Due the uncertainty principle,
$\nu_{min}<1/2$, so that the logarithmic negativity is given by

\begin{eqnarray}\label{negatividade}
 E_N =
       \begin{cases}

              \log \bigg(\sqrt{\frac{z_{0-}}{z_{0+}}}\bigg), & \mbox{if } z_{0+} \leq z_{0-} \mbox{;} \\ \\

              \log \bigg(\sqrt{\frac{z_{0+}}{z_{0-}}}\bigg), & \mbox{otherwise.}

       \end{cases}
\end{eqnarray}
which is propagation-distance-independent. We observe in Eq.
(\ref{negatividade}) that the entanglement measured by the
logarithmic negativity can be modified by changing the Rayleigh
length $z_{0+}$ since $z_{0-}$ is fixed by the laser pump
properties. On the other hand, the double Gaussian biphoton
wavefunction approximation shows no entanglement for
$z_{0+}=z_{0-}$. In the analysis of Ref. \cite{Matthew}, which uses
the Schmidt number, the entanglement is Rayleigh-length-dependent and
the case $z_{0+}=z_{0-}$ implies no entanglement within the spatial
phase of entangled photon pairs. In \cite{Matthew} the authors compute the
Schmidt number through their Eq. (6) for the double Gaussian wave
function, which, using our wave function Eq. (\ref{psifp}), can be
written as
\begin{equation}\label{schmidt}
\begin{split}
K_{dG}= &\bigg( \frac{w_+}{w_-}+ \frac{w_-}{w_+} \bigg)^2+k_0^2
w^2_+ w^2_-\bigg(\frac{1}{r_-}-\frac{1}{r_+}\bigg)^2 \\ =&
\bigg(\sqrt{\frac{z_{0-}}{z_{0+}}}+\sqrt{\frac{z_{0+}}{z_{0-}}}\bigg)^2,
\end{split}
\end{equation}
where $w_{\pm}$ and $r_{\pm}$ are given by Eq. (\ref{b}). From Eqs.
(\ref{negatividade}) and (\ref{schmidt}) we obtain
$E_N=\log(\sqrt{K_{dG}})$ for $z_{0+}/z_{0-}\ll1$ and
$z_{0+}/z_{0-}\gg1$. In Fig. \ref{schmidtfig} we compare the
logarithmic negativity with the logarithm of the root square of the
Schmidt number. In Fig. \ref{schmidtfig}a we plot the logarithmic
negativity and the logarithm of the root square of the Schmidt
number as a function of $z_{0+}/z_{0-}\leq1$ and in Fig.
\ref{schmidtfig}b we plot these quantities for $z_{0+}/z_{0-}\geq1$.
We observe an agreement of these quantities in the limits
$z_{0+}/z_{0-}\ll1$ and $z_{0+}/z_{0-}\gg1$.

\begin{figure}[htp]
\centering
\includegraphics[height=4cm,width=5.0 cm]{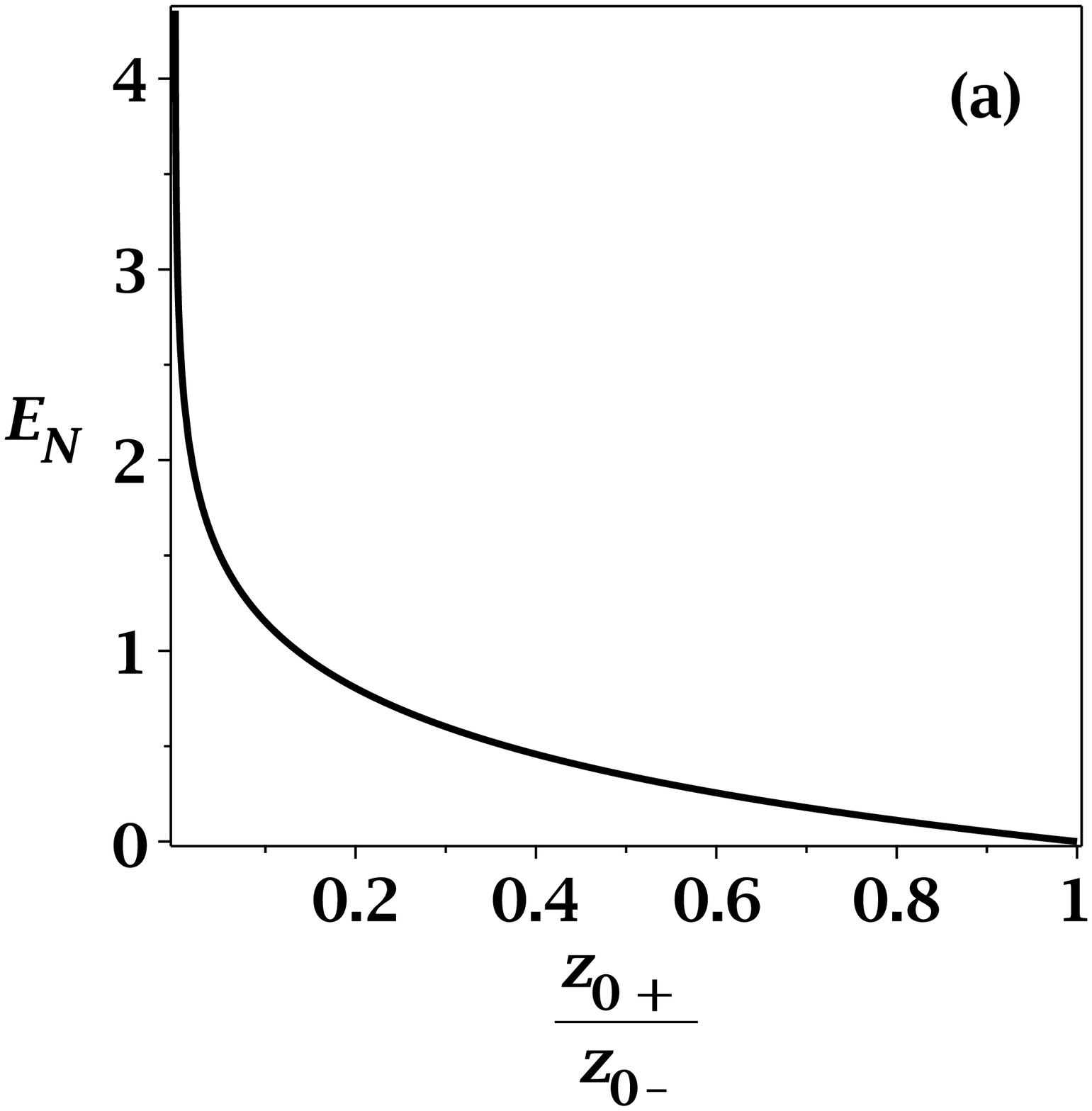}
\includegraphics[height=4cm,width=5.0 cm]{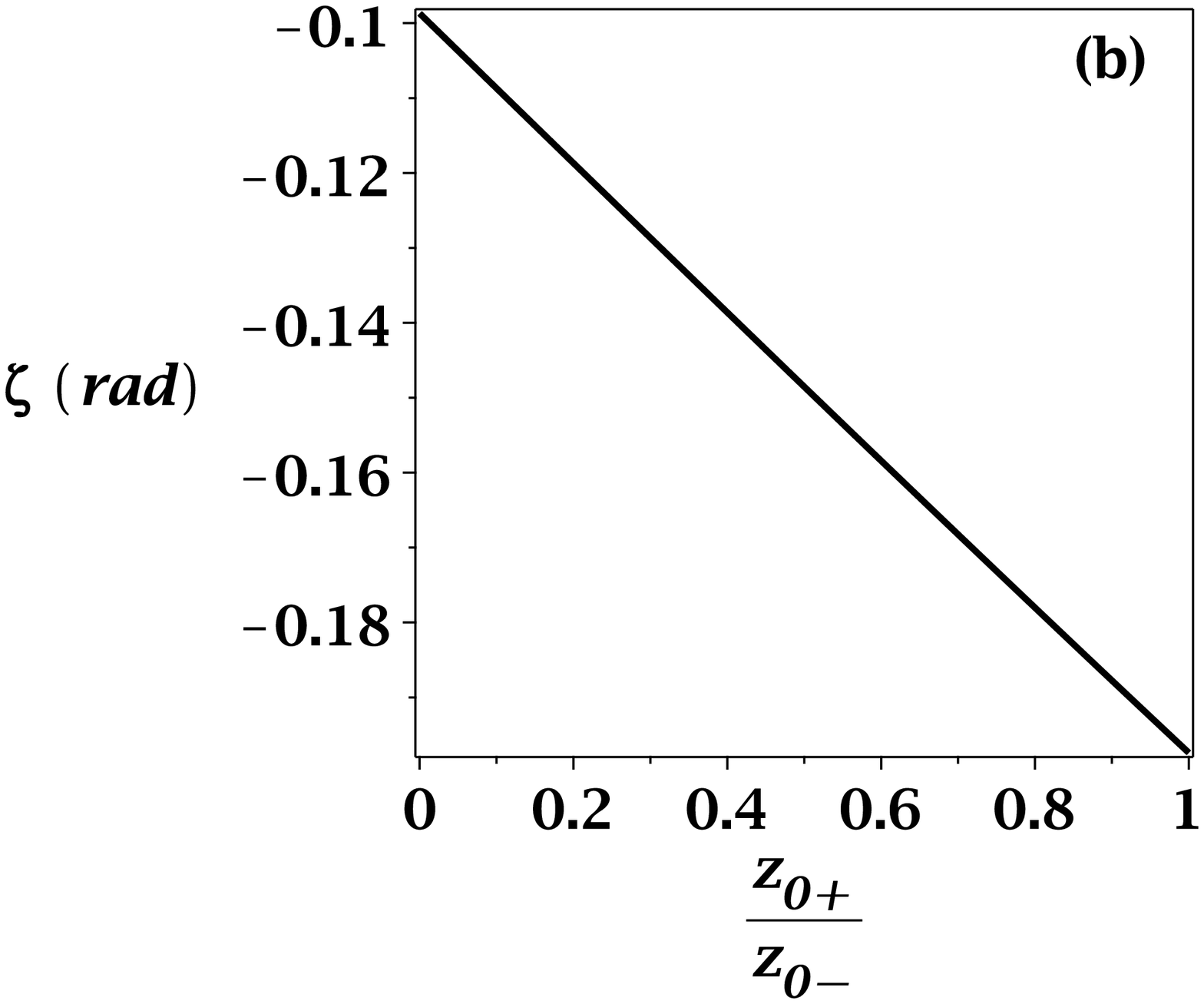}
\includegraphics[height=4cm,width=5.0 cm]{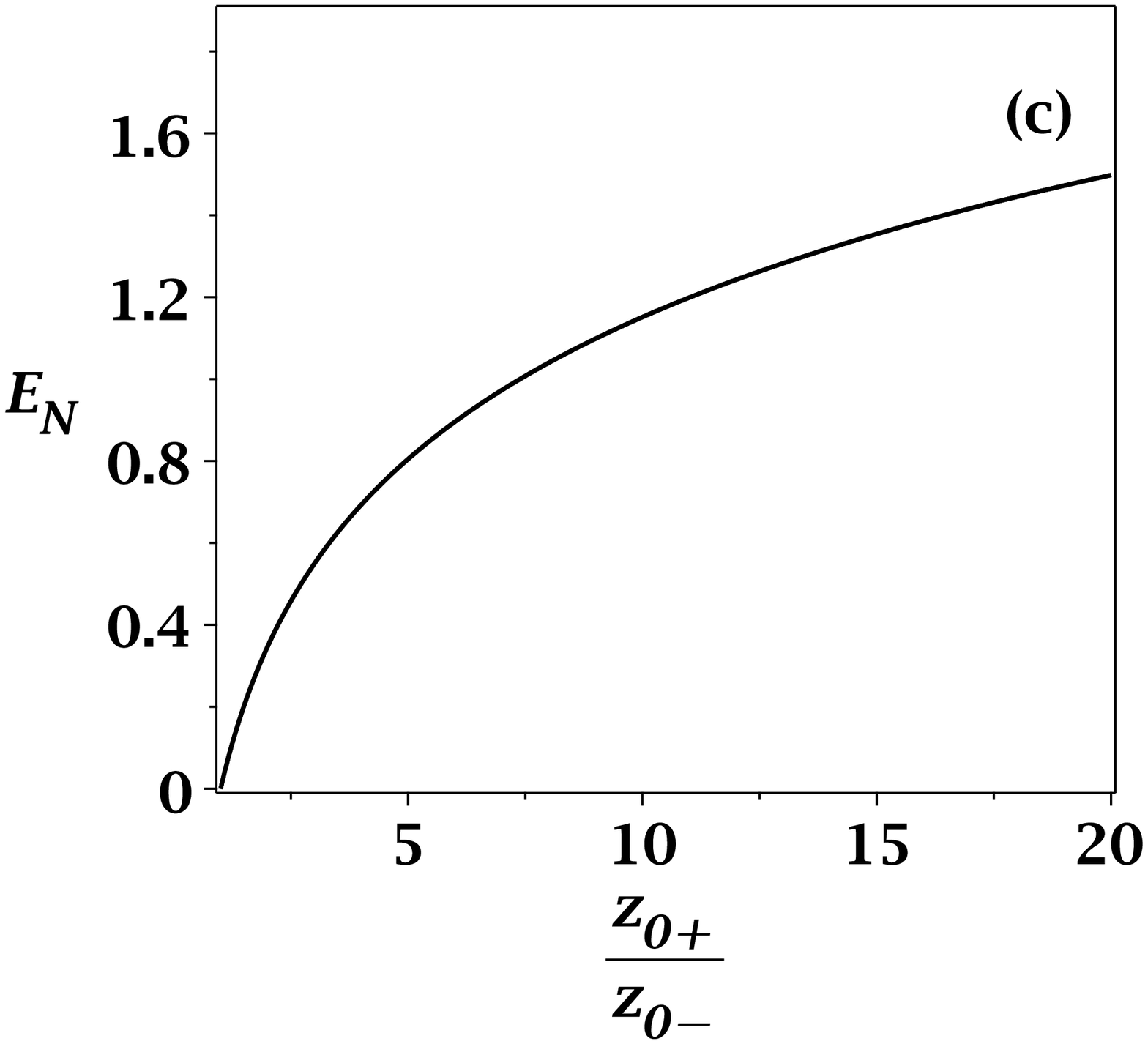}
\includegraphics[height=4cm,width=5.0 cm]{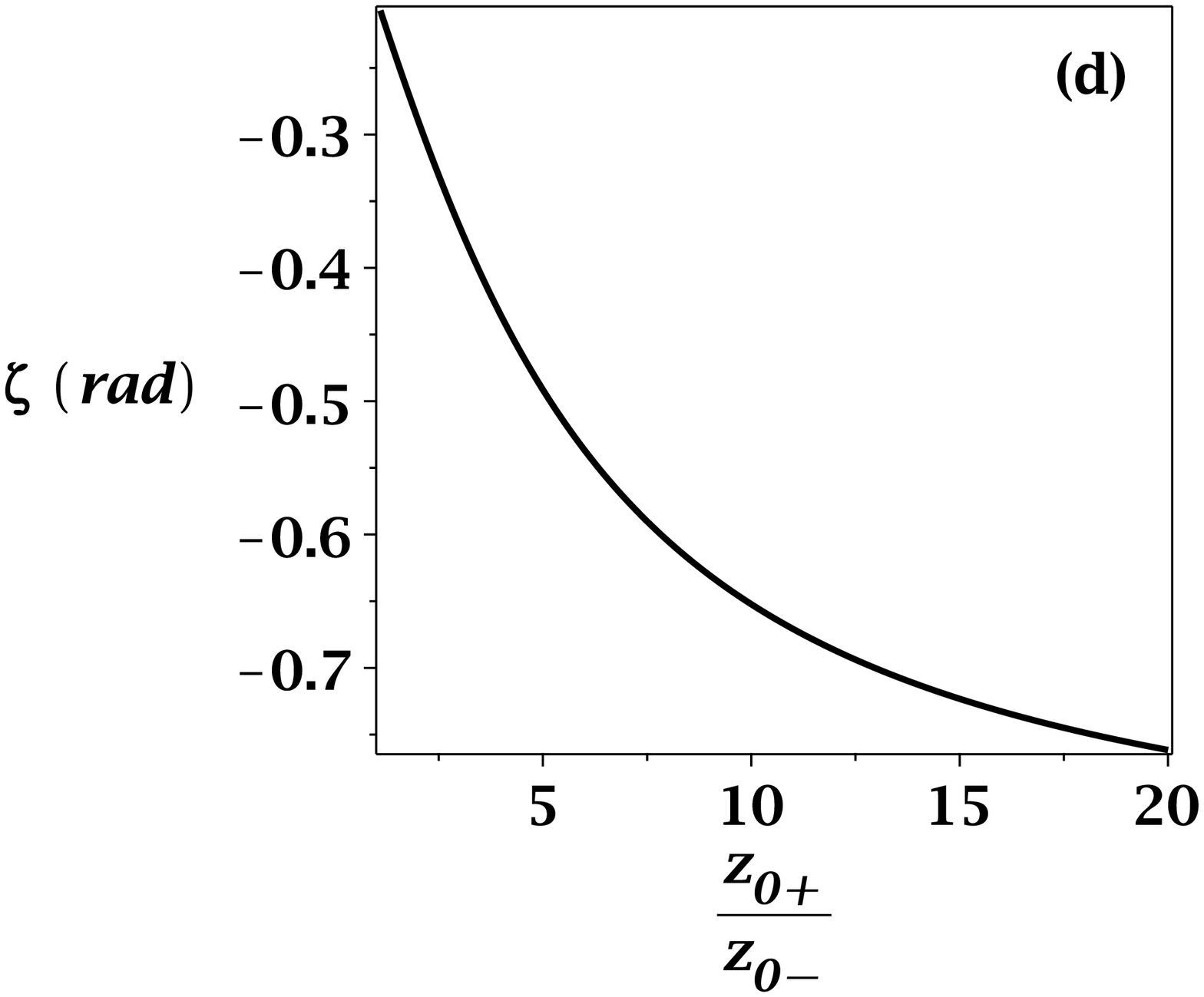}
\caption{(a) Logarithmic negativity and (b) Biphoton Gouy phase for
the Rayleigh range $z_{0+}/z_{0-}\leq1$, (c) Logarithmic negativity
and (d) Biphoton Gouy phase for the Rayleigh range
$z_{0+}/z_{0-}\geq1$, with $z_{0-}=1.2\;\mathrm{mm}$ and
$z=20\;\mathrm{mm}$. The logarithmic negativity varies appreciably
in both intervals whereas the Gouy phase variation is appreciable
only for the interval $z_{0+}/z_{0-}\geq1$ in which $z_{0+}$ tends
to $z$.} \label{gouy_z+}
\end{figure}

As the Gouy phase is a function of the Rayleigh length $z_{0+}$ and
$z_{0-}$, one can measure this longitudinal phase as a function of
the entanglement by changing the Rayleigh length $z_{0+}$ and fixing
the parameters $z$ and  $z_{0-}$ -- see Eq. (\ref{mu1mu2}). To observe
the behavior of logarithmic negativity and the Gouy phase as a
function of $z_{0+}$, we plot these quantities in Fig. \ref{gouy_z+}.
We consider $z_{0-}=1.2\;\mathrm{mm}$ and the longitudinal position
$z=20\;\mathrm{mm}$. In Fig. \ref{gouy_z+}a we plot the logarithmic
negativity and in Fig. \ref{gouy_z+}b the Gouy phase as a function
of $z_{0+}/z_{0-}\leq1$. In Fig. \ref{gouy_z+}c we plot the
logarithmic negativity and in Fig. \ref{gouy_z+}d the Gouy phase as
a function of $z_{0+}/z_{0-}\geq1$.

We observe that the logarithmic negativity suffers a large variation
for $z_{0+}/z_{0-}\leq1$ whereas the Gouy phase does not vary significantly. However, for $z_{0+}/z_{0-}\geq1$ the Gouy phase changes
appreciably. It is known that the Gouy phase varies the most within
the Rayleigh length. Therefore, the Gouy phase variation as a
function of $z_{0+}$ in the position $z=20\;\mathrm{mm}$ will be
small if $z\gg z_{0+}$ (which occurs for
$z_{0+}\leq1.2\;\mathrm{mm}$, i.e., for $z_{0+}/z_{0-}\leq1$) and
appreciable if $z_{0+}$ is of the order of $z$ (which occurs for
$z_{0+}\geq1.2\;\mathrm{mm}$, i.e., for $z_{0+}/z_{0-}\geq1$). In
the next section we will consider the two-dimensional propagation
through a thin lens which enables us to adjust existing experimental
data for the biphoton Gouy phase as a function of the shifted
Rayleigh length.

\section{Agreement with existing experimental data}
In Ref. \cite{Kawase} the authors showed for the first time the
relation between the Gouy phase and the quantum correlations of the
twin photons generated by parametric down conversion. Then, they
measured the coincidence count rates to experimentally obtain the
Gouy phase as a function of the position of the beam waist. In this
section we compare the biphoton Gouy phase with the experimental
data obtained in Ref. \cite{Kawase}. In that experiment they
considered as the pump a continuous wave (CW) argon-ion laser of
wavelength $\lambda_p=351\;\mathrm{mm}$ and power
$P=60\;\mathrm{mW}$, which was focused by a lens of focal distance
$f=900\;\mathrm{mm}$ to the beam radius $w_p=178\;\mathrm{\mu m}$ in
a BBO crystal of type I, which produces signal and idler photon
beams with the same wavelength $\lambda=702\;\mathrm{nm}$. Also,
they used lenses of focal distance $f=200\;\mathrm{mm}$ in the paths
of the signal and idler beams. Therefore, by changing the position
of the lens in the signal path (which corresponds to changing the
position of its beam waist) while scanning with a two-dimensional
hologram the idler path they were able to measure the coincidence
counting rates in different positions of the signal beam waist.
Then, by observing that the position of the maximum and minimum
coincidences becomes rotated by a phase that includes the Gouy phase
difference of the modes $LG_{00}$ and $LG_{0-1}$, they could relate
the quantum correlation with the Gouy phase.

In order to analyze the experimental data of Ref. \cite{Kawase} we
need to focus the biphoton wavefunction. Then, by considering a thin
lens approximation, and focal length $f$, the focused biphoton
wavefunction is given by

\begin{equation}
\begin{split}
\Psi(r,q,z,z')=& \int_{r',q'} K_r(r, r';z+z',z) K_q(q,q';z+z',z)
f(r', q') \psi(r', q',z),
\end{split}
\end{equation}
where the propagators $K_r$ and $K_q$ are given by Eq.
(\ref{propagador}), the state $\psi(r',q',z)$ is written as Eq.
(\ref{psifp}) and the transmittance of a thin lens is given by
\cite{Saleh,stoler}
\begin{equation}
f(r',q')= \exp \bigg[ - \frac{i k}{2f}(r'^2+q'^2) \bigg].
\end{equation}

\begin{figure}[htp]
\centering
\includegraphics[height=5cm,width=6.0 cm]{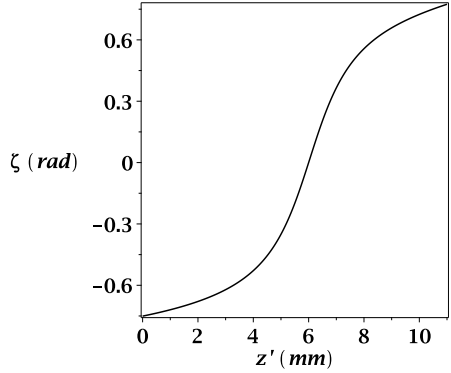}
\caption{Gouy phase for the focused biphoton wavefunction Eq.
(\ref{gouy_fit}) as a function of the position after the lens
$z^{\prime}$.} \label{gouy_zl}
\end{figure}

After some manipulations, we can write

\begin{equation}
\Psi(r,q,z,z')=\sqrt{\frac{2}{ \pi B_{+}
B_{-}}}\exp\left(-\frac{r^2}{B^2}\right)\exp\left(-\frac{q^2}{B_{-}^2}\right)\exp\left[\frac{ik_0
}{ c R_{+}}r^2+\frac{ik_0 }{ c R_{-}}q^2-i\zeta(z,z')\right],
\end{equation}
where

\begin{equation}\label{B}
B_{+}^2(z, z')=\frac{\left(
\frac{1}{w_{+}^2}\right)^2+k_0^2\left(\frac{1}{z'}+\frac{1}{c
r_{+}}-\frac{1}{2f}\right)^2}{\left(\frac{2\pi}{\lambda
z'}\right)^2\left( \frac{1}{w_{+}^2}\right)},
\end{equation}

\begin{equation}
B_{-}^2(z, z')=\frac{\left(
\frac{1}{w_{-}^2}\right)^2+k_0^2\left(\frac{1}{z'}+\frac{1}{c
r_{-}}-\frac{1}{2f}\right)^2}{\left(\frac{2\pi}{\lambda
z'}\right)^2\left( \frac{1}{w_{-}^2}\right)},
\end{equation}

\begin{equation}
R_{+}(z,z')=
\frac{\left(\frac{1}{w_{+}^2}\right)^2+k_0^2\left(\frac{1}{z'}+\frac{1}{c
r_+}-\frac{1}{2f}\right)^2}{\frac{c}{z'w_{+}^2}\left(1
+\frac{1}{\Omega^2}\left(\frac{z}{z'} + \frac{z}{c
r_+}\right)\right)-\frac{\pi}{\lambda f}},
\end{equation}

\begin{equation}
R_{-}(z,z')=
\frac{\left(\frac{1}{w_{-}^2}\right)^2+k_0^2\left(\frac{1}{z'}+\frac{1}{c
r_-}-\frac{1}{2f}\right)^2}{\frac{c}{z'w_{-}^2}\left(1
+\frac{1}{\sigma^2}\left(\frac{z}{z'} + \frac{z}{c
r_-}\right)\right)-\frac{\pi}{\lambda f}},
\end{equation}
and
\begin{equation}\label{gouy_fit}
\zeta(z,z')= \frac{1}{2} \arctan \Bigg\lbrace
\frac{\big(\frac{z}{1-z'/2
f}+z'\big)\big(\frac{1}{z_{0+}}+\frac{1}{z_{0-}}
\big)}{1-\frac{1}{z_{0+}z_{0-}} \big(\frac{z}{1-z'/2 f}+z'\big)^2}
\Bigg\rbrace.
\end{equation}
Now, the parameters of the wavefunction are dependent on the focal
distance $f$. By the analogy with the focused classical Gaussian
beam, $B_{\pm}(z,z^{\prime})$ is the corresponding beam width,
$R_{\pm}(z,z^{\prime})$ is the corresponding radius of curvature of
the wavefronts and $\zeta(z,z^{\prime})$ is the corresponding Gouy
phase. In the limit $f\rightarrow\infty$ we recover the parameters
of the biphoton wavefunction in Eq. (\ref{psifp}) for the
propagation $z+z^{\prime}$. In Fig. \ref{gouy_zl} we plot the Gouy
phase for the focused biphoton wavefunction Eq. (\ref{gouy_fit}) as
a function of the position after the lens $z^{\prime}$. We consider
the following parameters $z_{0+}=z_{0-}=1.2\;\mathrm{mm}$,
$f=3.0\;\mathrm{mm}$ and $z=7.0\;\mathrm{mm}$. As we can observe the
phase is null for $z^{\prime}=2f=6.0\;\mathrm{mm}$. Again the
maximum Gouy phase variation is $\pi/2$ as we have considered the
one-dimensional focalization.

\begin{figure}[htp]
\centering
\includegraphics[height=5cm,width=6.0 cm]{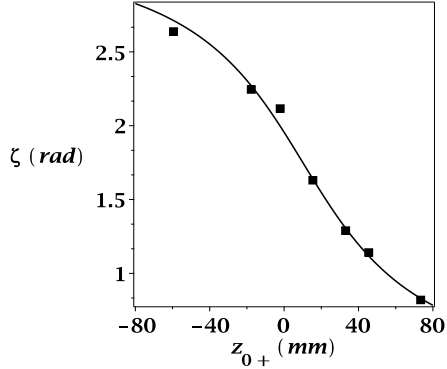}
\caption{Gouy phase as a function of $z_{0+}$. The squares
represents the experimental data from \cite{Kawase} and the solid
line represents the fitting result Eq. (\ref{gouy2D2}).
\label{kawase2}}
\end{figure}

Now, in order to use the biphoton Gouy phase to fit the experimental
data of Ref. \cite{Kawase} we need to rewrite Eq. (\ref{gouy_fit})
to include the two-dimensional propagation through a thin lens which
transforms it to
\begin{eqnarray}\label{gouy2D2}
\zeta(z_{0+})=  \zeta_0+ \arctan \Bigg\lbrace
\frac{\big(\frac{z}{1-z'/2
f}+z'\big)\big(\frac{1}{(z'_{0+}-z_f)}+\frac{1}{z_{0-}}
\big)}{1-\frac{1}{(z'_{0+}-z_f)z_{0-}} \big(\frac{z}{1-z'/2
f}+z'\big)^2} \Bigg\rbrace,
\end{eqnarray}
where $\zeta_0$ is a reference angle and $z_f$ an adjust parameter.
In Fig. \ref{kawase2} we show the Gouy phase as a function of the
Rayleigh range $z_{0+}$ shifted by an offset distance $z_{offset}$.
The negative values appearing for $z_{0+}$ in the horizontal axis is
a consequence of the shift by $z_{offset}$. The squares represent
the experimental data from \cite{Kawase} and the solid line
represents the fitting result by Eq. (\ref{gouy2D2}). As discussed
before the two-photon entanglement is included in $z_{0+}$.
In order to adjust the experimental data of Fig. 4 of Ref.
\cite{Kawase} with Eq. (\ref{gouy2D2}) we used the \textit{Maple}
software which produces the following values of parameters: biphoton
wavelength $\lambda=702\;\mathrm{nm}$, laser pump wavelength
$\lambda_p=351.1\;\mathrm{nm}$ and the crystal length
$L_p=7.0\;\mathrm{mm}$. This enables us to obtain
$\sigma=\sqrt{\frac{L_p \lambda_p}{6 \pi}}=11.4\;\mathrm{\mu m}$,
$z_{0-}=k_0\sigma^{2}=1.2\;\mathrm{mm}$, $f=200\;\mathrm{mm}$,
$z=500\;\mathrm{mm}$, $z^{\prime}=1465.3\;\mathrm{mm}$,
$\zeta_0=1.68\;\mathrm{rad}$ and $z_f=7.15\;\mathrm{mm}$. Because of
some effect of the experimental arrangement, such as that produced
by the hologram, we need to include a parameter $z_f$ in Eq.
(\ref{gouy2D2}) in order to adjust the experimental data. The
reasonable agreement between theory and experimental data on the
Gouy phase indicates the double Gaussian wavefunction  is a valid
approximate description of two correlated photons generated by
type-I SPDC.

\begin{figure}[htp]
\centering
\includegraphics[height=5cm,width=6.0 cm]{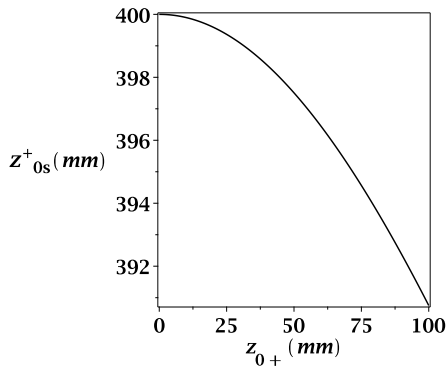}
\caption{Beam waist position $z^{+}_{0s}$ as function of the
Rayleigh range $z_{0+}$.\label{zmin}}
\end{figure}

In Ref. \cite{Kawase} the Gouy phase was obtained by changing the
position of the beam width $z_{0s}$. Here, we adjust the Gouy phase
by changing the Rayleigh length $z_{0+}$ instead of $z_{0s}$. Now,
we will show that these two quantities are related. The beam waist
position $z^{+}_{0s}$ after a thin lens can be obtained from Eq.
(\ref{B}) and written as
\begin{equation}
z^{+}_{0s}= \frac{2 c k_0^2  w_+^4 r_+ (c r_+ - 2 f)f}{ k_0^2 w_+^4
(c r_+ - 2 f)^2+ 4 f^2 c^2 r_+^2},
\end{equation}
where $w_+$ and $r_+$ are given by Eq. (\ref{b}), $k_0$ is the
wavenumber Eq. (\ref{r}), $c$ is the speed of light and  $f$ is
the focal length. This quantity is $z_{0+}$-dependent trough the
parameters $w_+$ and $r_+$. In order to observe the behavior of the
beam waist position $z^{+}_{0s}$ as a function of the Rayleigh
length $z_{0+}$, we plot it in Fig. \ref{zmin}. This plot shows that
the beam waist position varies with the Rayleigh length. We consider
the same parameters of Fig. \ref{kawase2}.
Therefore, this relation is the reason why one can also plot the
experimental data of Ref. \cite{Kawase} as a function of the
Rayleigh length. In addition, although the authors used a
superposition of LG modes to observe the Gouy phase instead of a
Gaussian mode, they observed that the superposition is converted
into a Gaussian mode when the hologram is shifted and scanned to
change the phase between LG modes. As we can see the expression
found in Eq. (8) of Ref. \cite{Kawase} is characteristic of Gouy
phase for Gaussian beams.

\section{Concluding remarks}
We considered the time (or longitudinal distance) propagation of the
approximated double Gaussian wavefunction describing correlated
photons generated in a nonlinear crystal. We considered photons
generated in a type I-SPDC process, in which the twin photons have
the same wavelength. We found that the evolved wavefunction is
characterized by parameters similar to that of a classical Gaussian
beam, specially by a Gouy phase term. Next, we studied the twin
photon entanglement by calculating the covariance matrix and the
logarithmic negativity for the double Gaussian wavefunction at the
propagation distance. We observed that the Gouy is part of the
elements of the covariance matrix through the position momentum
covariance that develop with the propagation distance. Then, we
showed that the logarithmic negativity is a function of the Rayleigh
length and the biphoton Gouy phase can be obtained by changing the
entanglement through the Rayleigh length. We also compare the
logarithmic negativity with the Schmidt number and found that both
entanglement quantifiers are Rayleigh-length-dependent such that for
specific limits the first entanglement quantifier is the logarithm
of the root square of the second quantifier. Furthermore, we
considered an experiment performed with entangled photons generated
in a type-I SPDC process, in which the Gouy phase was measured as a
function of the signal beam waist position. By knowing that the beam
waist position and the Rayleigh range are related when a beam is
focused by a lens, we focused the double Gaussian biphoton
wavefunction by a thin lens and adjusted the experimental data as a
function of the Rayleigh range. We obtained a reasonable agreement
between the biphoton Gouy phase and the experimental data. This
agreement between theory and experiment indicates that the Gouy
phase of the approximated double Gaussian biphoton wavefunction can
be used as good approximation in exploring quantum correlations of
twin photons.

Our results show that the biphoton Gouy phase and the entanglement
are Rayleigh length dependents enabling us to connect these two
quantities. The Rayleigh length is focal spot dependent allowing to
interpret both quantities in the same physical origin, i.e., the
transverse spatial confinement. Also, it is known that these
quantities have geometrical features which is the reason why they
are spatial confinement dependent \cite{Simon}. Therefore, based in
the spatial confinement by slits, we are going to propose in a
future paper a way to measure the biphoton Gouy phase to obtain the
corresponding portion of entanglement correlations.
\section*{Acknowledgments}
The authors would like to thank Coordena\c c\~ao de Aperfei\c
coamento de Pessoal de N\'{\i}vel Superior (CAPES) and Conselho
Nacional de Desenvolvimento Cient\'{\i}fico e Tecnol\'ogico (CNPq)
for the financial support. J.B.A thanks CNPq for the Grant No.
150190/2019-0. I. G. da Paz thanks CNPq for the Grant No.
307942/2019-8. M.S. thanks CNPq for the Grant No. 303482/2017, and
Funda\c c\~ao de Amparo \`A  Pesquisa do Estado de S\~ao Paulo
(FAPESP) for the Grant No. 2018/05948-6.  B.H. acknowledges  support
by CFisUC and  Funda\c c\~ao  para a   Ci\^{e}ncia   e   Tecnologia
(FCT) through the project UID/FIS/04564/2016.


\end{document}